\documentstyle[12pt]{article}
\begin{document}
\title{Origin of Life in the Universe}
\author{B.G. Sidharth$^*$\\
B.M. Birla Science Centre, Adarsh Nagar, Hyderabad - 500 063 (India)}
\date{}
\maketitle
\footnotetext{$^*$Email:birlasc@hd1.vsnl.net.in; birlard@ap.nic.in}
\begin{abstract}
We briefly examine the question of the origin of life, both terrestrial
and extraterrestrial, in the light of latest findings and point out that
the data is consistent with a dual mode origin: Some of the ingredients
including possibly sugars being brought to earth from outer space which
together with other ingredients for example proteins
already present on the earth would lead to the formation of life.
\end{abstract}
\section{Introduction}
We all know broadly what life is, but even today it is difficult to give a
precise definition. An extreme example from the inanimate world of something
which resembles life is the case of certain crystals which can replicate or
reproduce. But life means a lot more - for example growth and metabolism and
all this within the context of an interaction with the environment. An
extreme example from the  low end of the animate world is that of a virus.
As of today it is very difficult to precisely bridge the apparently narrow
gap between the inanimate and the animate. Apart from these nuances a
question that arises is, how did life originate in the first place?
\section{The Origins of Life}
In the above context, the discovery of Friedrich Wohler in the nineteenth
Century that Urea which is so interlinked with living creatures could be
made in the laboratory with inanimate chemical reactions was a crucial first
step. Was it then possible that life itself could have evolved on the earth,
through a fortuitous combination of circumstances combined with ordinary
chemical reactions?\\
Actually, broadly three substances are crucial for living creatures. These
are carbohydrates, fats and proteins. Carbohydrates are basically carbon
compounds with water. These include sugars which can be found for example
in fruits, honey and even certain vegetables.  They play a role in the
storage of energy. Fats are long chain molecules, which include waxes and
oils. They form (nearly insoluble) suspensions in water and coalesce to form
distinct units. This helps in the formation of cells\cite{r1}. Finally,
proteins are highly complex molecules containing linked amino-acids,
containing compounds of Nitrogen and Hydrogen. The three together form the
cells of living creatures, including plants and micro organisms.\\
More specifically living organisms consist of cells. Each cell has a small
nucleus surrounded by a larger  block of material, the Cytoplasm. The
nucleus consists of long molecular threads or chromosomes, which issue
instructions to the Cytoplasm to produce other complex organic molecules
required for life processes like growth and survival.
The chromosomes themselves are made up of long molecular chains or
nucleic acids, which themselves are made up of four different kinds of
smaller units called nucleotides, for example the DNA. The Cytoplasm on
the other hand consists of long molecular chains, proteins, which themselves
are made up of twenty different combinations of smaller molecular chains
called the amino acids. The nucleotides and the amino acids are the basic
building blocks. They are made up of about thirty electrically bound atoms
of Hydrogen, Carbon, Nitrogen and Oxygen.\\
Scientists realised that for life to form on the earth, first amino acids,
sugars and other complex molecules needed to be present. These are born out
of various chemical reactions. Next these materials had to become proteins,
and what we call today nucleic acids. It was believed that the relatively
warm seas and oceans of the earth were the container for this chemistry,
in which living cells finally developed. These in turn would use sun light
to split water molecules, and would also capture carbon dioxide to
synthesise glucose. In the process they would release Oxygen into the
atmosphere.\\
The all important question was, could the amino acids and sugars have been
formed on the earth? The answer came in the 1950s when Harold Urey and
Stanley Miller of the University of Chicago subjected various gases such as
Ammonia, Methane, Hydrogen and Water vapour to lightning flashes\cite{r2}.
This resulted in the building up of four of the twenty common amino acids
in proteins. These experiments have since been refined and confirmed. So it
is generally accepted that a few billion years ago given the right
conditions, life evolved from the seas and oceans on the earth.\\
However, in the nineteenth Century itself, there was speculation that ready
made life came down to earth from outer space. There were a few variants of
this theme. For example life could have been brought down to earth by radio
waves. This theory was called Radiopanspermia. Another line of thinking
was that life could have been transported from outer space by Meteorites, a
theory that went by the name, lithopanspermia.\\
In more recent times the extraterrestrial origin of life has received some
support, though it remains a minority view. One interesting rationale for
this view is that evidence from fossilised rocks in Greenland points to
microscopic forms of life as early as about a billion years after the
formation of the earth. It could be argued that this time span is much
too short to provide the rather special circumstances for the formation of
living organisms.\\
On the other hand it is interesting that complex molecules have been found
in interstellar space, for example in the cool dust clouds of the Orion
Nebula and in the constellation of Sagittarius. Observations with telescopes,
spectrascopes, radio telescopes and even orbiting observatories have
confirmed the presence of molecules like Methyl Cyanide, Water vapour,
Formal Dehyde, Methyl Alcohol and even the potable Ethyl Alcohol. Clearly
there are certain organic molecules in the cool dust clouds spread across
outer space. In any case the vast life span of the universe and its great
variety in different parts provides a stage where it is plausible that
primitive life forms could have been synthesised.\\
Based on these considerations Sir Fred Hoyle and Prof. Chandra
Wickramasinghe have proposed\cite{r3} that life has been transported to the
earth from outer space. Such life forms could have been delivered to the
earth through debris of comets. These comets which inhabit the cool and
dark regions of the Solar System do contain the building blocks of life.
There is increasing evidence through spectroscopic, space craft and even
laboratory examinations of  debris fallen on the earth, that the frozen
dirty ice balls, as comets have been characterised contain not just
molecular compounds of  Carbon, Hydrogen, Nitrogen and Oxygen, but also
even more complicated sugar related substances. Latest studies at NASA's
Ames Research Centre point to the presence of polyhydroxilated compounds.\\
Interestingly a few years ago samples of a Meteorite, found in Antarctica
and believed to have been ejected from Mars some 3.5 billion years ago,
were independently examined by NASA and a team from England. Traces of
fossilised micro organisms were reported.\\
This conclusion if finally reconfirmed and accepted would have exciting
implications: there would have been life on Mars some 3.5 billion years ago,
at about the same time that life is believed to have surfaced on the earth.
This would then point to a common origin - for example through
transportation by cometary debris.\\
In any case, the latest findings of sugar related compounds and other
complex molecules in meteoritic and cometary objects would also support
a suggestion by the author some years ago, that infact the formation of
life on the earth could have been triggered by a dual process: some of these
molecules possibly sugars, but not yet living organisms being transported
to the earth, and then further biochemical reactions taking place on the
earth itself with for example proteins \cite{r4}. Such a hybrid view for the origin
of life is consistent with observation and experiment.
\section{Life Elsewhere}
Going beyond the question of life on earth itself, it is generally
accepted that there could be millions, if not billions of other candidates
in the universe for harbouring life. Life as we know it can survive not in
the hot interior of stars, nor in the frozen outer reaches of interstellar
space, but  more likely on planet like objects attached to stars. Such
objects would be neither too hot nor too cold, and they could have gaseous atmospheres,
all of which could support life. In recent years there has been direct and
indirect evidence for such planetary systems, and most recently it appears
that atleast one of these extra solar planets even has an atmosphere. We are
severely handicapped with the very limited reach of our observations.
Hopefully the European Space Agency's spacecraft GAIA\cite{r5},
which could be launched in 2010-11 would scan hundreds of samples
and would provide more definitive answers.

\end{document}